\documentclass[a4paper, amsfonts, amssymb, amsmath, reprint, showkeys, nofootinbib, twoside]{revtex4-1}
\usepackage{graphicx}  % needed for figures
\usepackage{dcolumn}   % needed for some tables
\usepackage{bm}        % for math
\usepackage{amssymb}   % for math
\usepackage{dutchcal}
\usepackage[toc,page]{appendix}
\usepackage{amsmath}
\usepackage{braket}
\usepackage{mathtools}
\begin{document}

\title{Canonical transformations applied to the non-free Landau electron}
\author{Jorge A. Lizarraga}
\email{jorge\_lizarraga@icf.unam.mx\\}
\affiliation{Instituto de Ciencias F\'isicas, Universidad Nacional Aut\'onoma de M\'exico, 62210, Cuernavaca, M\'exico}

\date{\today}

\begin{abstract}
The method previously used to solve Schr\"odinger equation by a unitary transformation for a electron under the influence of a constant magnetic field is used to obtain a non-free Landau electron wave function. The physical meaning of this wave function is discussed based on the conserved properties of the transformed Hamiltonian. 
\end{abstract}
\maketitle

\section{Introduction}
Previously, Boon and Seligman has shown how to recover Landau's solution for a particle under the effect of a constant magnetic field, described by the Landau's gauge, using a canonical transformation which correspond to a finite linear symplectic transformation in phase space \cite{boon1973canonical}, following the approach studied by Moshinsky and Quesne \cite{moshinsky1971linear,quesne1971canonical,moshinsky1972canonical}. This result has been called as {\it free Landau electron} due to the fact that the wave functions along one axis (depending on the gauge selected) is a plane wave function. However, this description turns out to be counterintuitive, since, classically talking when we restrict the system to the $x-y$ plane, under the influence of a magnetic field one spect to have circular movement which cannot be fully described by plane waves. In this work, I applied the same canonical transformation given on reference \cite{boon1973canonical} to show that there is a non-free Landau electron wave function for the same system which is characterized by oscillations in both axis of the plane $x-y$. Even though, whether at a first glance on the solution one could think that this wave function is a consequence of the degeneration of the system, it comes out that the apparition of this second term in the wave function has its origin on the conserved quantities of the Hamiltonian.

\section{Canonical transformation}
The system Hamiltonian is written as 
\begin{equation}
{\hat H}=\frac{1}{2m}\left({\bf {\hat p}}+\frac{e}{c}\bf{A}\right)^{2},
\end{equation}
where $e>0$ is the electron charge, $m$ is its mass, $c$ the speed of light, ${\bf {\hat p}}=-i\hbar\nabla$ is the particle momentum and ${\bf A}=(A_{x},A_{y},A_{z})$ is the vector potential such that ${\bf B}=\nabla\times{\bf A}=B{\hat k}$. Hence, the gauge is selected as
\begin{equation}
{\bf A}=(-By,0,0),
\end{equation}
which is known as Landau's gauge. Since the interesting part of the particle dynamics is happening inside the plane $x-y$, due to the fact the coordinate along the $z$-axis is unaffected by the magnetic field, defining $\omega_{c}=eB/mc$, the Hamiltonian has the following form
\begin{equation}\label{H}
{\hat H}=\frac{1}{2m}({\hat p}_{x}-m\omega_{c}y)^{2}+\frac{1}{2m}{\hat p}_{y}^{2},
\end{equation}
the method on reference \cite{de2021solution} can be used to calculate the eigenvalues without having to calculate the eigenfunctions (this method works even without having to specify the gauge used), being the spectrum the Landau levels $E_{n}=\hbar\omega_{c}\left(n+1/2\right)$ where $n\in\mathbb{N}$. Now, we used the same linear transformation used on reference \cite{boon1973canonical} which redefines the canonical (operator) variables  $(x,y,{\hat p}_{x},{\hat p}_{y})$ to (operator) variable $(Q,{\overline Q},P,{\overline P})$
\begin{equation}\label{Q}
Q=-\frac{1}{\beta}({\hat p}_{x}-\beta y),
\end{equation}
\begin{equation}\label{Q2}
{\overline Q}=-\frac{1}{\beta}({\hat p}_{y}-\beta x),
\end{equation}
\begin{equation}\label{P}
P={\hat p}_{y},
\end{equation}
and
\begin{equation}\label{P2}
{\overline P}={\hat p}_{x},
\end{equation}
where $\beta=m\omega_{c}$. The
context should always be clear.Then the following commutation relations can be calculated 
\begin{equation}\label{iden1}
[Q,P]=[\overline{Q},\overline{P}]=i\hbar
\end{equation}
and
\begin{equation}\label{iden2}
[Q,\overline{Q}]=[P,\overline{P}]=[Q,\overline{P}]=[\overline{Q},P]=0,
\end{equation}
warranting that the new variables are canonical ones (expression (2.1) in reference \cite{moshinsky1971linear}). The Hamiltonian Eq. (\ref{H}) is rewritten with this coordinates as
\begin{equation}\label{HQP}
{\hat H}=\frac{1}{2m}\beta^{2}Q^{2}+\frac{1}{2m}P^{2}.
\end{equation}
To recover Landau's solution \cite{landau2013quantum} in therms of the variables $x$ and $y$ the Moshinsky and Quesne formula was used \cite{moshinsky1971linear}
\begin{equation}\label{}
\begin{aligned}
\psi(x,y)=&\frac{\beta}{2\pi}\int\limits_{-\infty}^{\infty}\int\limits_{-\infty}^{\infty}\exp\left[i\frac{\beta}{\hbar}(Q\overline{Q}+xy-xQ-y\overline{Q})\right]\\
&\times\psi(Q,\overline{Q})dQd\overline{Q},
\end{aligned}
\end{equation}
where $\psi(Q,\overline{Q})$ is an arbitrary function that depends on the variables $Q$ and $\overline{Q}$. Due to the independence of the Hamiltonian respect the variable $\overline{Q}$ the solution can be written as 
\begin{equation}\label{}
\psi(Q,\overline{Q})=\Psi_{n}((\beta/\hbar)^{1/2}Q)\phi(\overline{Q}),
\end{equation}
where $\Psi_{n}$ is the solution of the Harmonic oscillator equation and $\phi=\exp\left(i\frac{k}{\hbar}\overline{Q}\right)$ \cite{boon1973canonical}. Under this consideration, the solution obtained is 
\begin{equation}\label{landauSol}
\psi_{n,k}(x,y)=\exp\left(i\frac{k}{\hbar}x\right)\Psi_{n}\left(\sqrt{\frac{\beta}{\hbar}}\left(y-\frac{k}{\beta}\right)\right).
\end{equation}
Hence, Landau's solution was obtained. However, due to the arbitrariness of the function $\psi(Q,\overline{Q})$, one can define the expression 
%\begin{equation}\label{}
%\begin{aligned}
%\psi(Q,\overline{Q})=&\Psi_{n}((\beta/\hbar)^{1/2}Q)\exp\left(i\frac{k}{\hbar}\overline{Q}\right)+\\
%&\Psi_{n}((\beta/\hbar)^{1/2}Q)\delta\left(\overline{Q}-\frac{k'}{\beta}\right),
%\end{aligned}
%\end{equation}

\begin{equation}\label{}
\psi=\Psi_{n}((\beta/\hbar)^{1/2}Q)\left(\exp\left(i\frac{k}{\hbar}\overline{Q}\right)+\delta\left(\overline{Q}-\frac{k'}{\beta}\right)\right),
\end{equation}
where $\delta$ is the Dirac delta function. Of course, the first term of the above definition will give the expression Eq. (\ref{landauSol}). On the other hand, the second term will lead us to a harmonic oscillator solution along the $x$-axis times a phase, that is 
\begin{equation}\label{nonfree}
\begin{aligned}
\psi_{n,k,k'}(x,y)=\exp\left(i\frac{k}{\hbar}x\right)\Psi_{n}\left(\sqrt{\frac{\beta}{\hbar}}\left(y-\frac{k}{\beta}\right)\right)+\\\\
\exp\left(i\frac{\beta}{\hbar}\left(x-\frac{k'}{\beta}\right)y\right)\Psi_{n}\left(\sqrt{\frac{\beta}{\hbar}}\left(x-\frac{k'}{\beta}\right)\right).
\end{aligned}
\end{equation}
This is the non-free Landau electron wave function, due the fact that the particle description along the $x$-axis is not longer a plane wave alone, but it also present oscillations along it.

\section{Physical interpretation}
To understand the physical meaning of the expression Eq. (\ref{nonfree}) it is useful to analyze the conserved quantities of the Hamiltonian Eq. (\ref{HQP}). In Heisenberg picture, the evolution of a time-independent operator is given by the expression  
\begin{equation}
\frac{d{\hat f}}{dt}=\frac{1}{i\hbar}[{\hat f},{\hat H}],
\end{equation}
using Eq. (\ref{HQP}) and the identities Eq. (\ref{iden1}) and Eq.(\ref{iden2}) one can calculate the following variations 
\begin{equation}\label{var1}
\frac{d Q}{dt}=\frac{i\hbar}{m}P,
\end{equation}
\begin{equation}\label{var2}
\frac{d \overline{Q}}{dt}=0,
\end{equation}
\begin{equation}\label{var3}
\frac{d P}{dt}=-\frac{i\hbar}{m}\beta^{2}Q,
\end{equation}
and
\begin{equation}\label{var4}
\frac{d \overline{P}}{dt}=0.
\end{equation}
The above equalities tell us that this system has two conserved quantities, besides the Hamiltonian itself, Eq. (\ref{var2}) and Eq.(\ref{var4}), this is in fact expected since the Hamiltonian is independent of the quantities $(\overline{Q},\overline{P})$. This result tell us that 
\begin{equation}\label{dpxdt}
\frac{d {\hat p}_{x}}{dt}=0,
\end{equation}
and 
\begin{equation}\label{dpydt}
\frac{d }{dt}({\hat p}_{y}-\beta x)=0,
\end{equation}
are constant of motion. The first one of the above conserved operator is the one that was initially considered by Landau to obtain his solution \cite{landau2013quantum}, by solving the eigenvalue equation 
\begin{equation}
\overline{P}\psi=k\psi,
\end{equation}
and substituting it in the Hamiltonian, one gets the wave function Eq. (\ref{landauSol}). On the other hand, one can proceed similarly with the conserved operator $\overline{Q}$, writting down a eigenvalue expression 
\begin{equation}
\overline{Q}\psi=-\beta^{-1} k'\psi,
\end{equation}
substitute the result in the Hamiltonian and the solution one gets is the second term of the expression Eq. (\ref{nonfree}). One needs to mention that the operator Eq. (\ref{dpydt}) has been ignored so far.

An insight of the physical meaning of the conserved quantities can be obtained when one analyze the characteristics of the Lorentz force being applied to the particle due to the magnetic field. Such force is defined as 
\begin{equation}
{\bf F}=-\frac{e}{c}{\bf v}\times{\bf B},
\end{equation}
using it to express the coordinate representation of the force in the plane $x-y$, one can obtain the following conserved expressions 
\begin{equation}\label{Ncons1}
\frac{d}{dt}\left(m\frac{dx}{dt}+\beta y\right)=0,
\end{equation}
and
\begin{equation}\label{Ncons2}
\frac{d}{dt}\left(m\frac{dy}{dt}-\beta x\right)=0.
\end{equation}
In the Hamiltonian formalism the Newtonean momentum defined as ${\bf p}=m{\bf v}$ does not necessary coincides with the momentum defined by the Hamiltonian. In fact, using the identities Eq. (\ref{var1}),  Eq. (\ref{var2}),  Eq. (\ref{var3}) and Eq.(\ref{var4}) is not difficult to prove that, for this particular gauge selection, one has that 
\begin{equation}
p_{x}=m\frac{dx}{dt}+\beta y,
\end{equation}
and 
\begin{equation}
p_{y}=m\frac{dy}{dt},
\end{equation}
Hence, the conserved expressions Eq. (\ref{Ncons1}) and Eq.(\ref{Ncons2}) automatically becomes Eq. (\ref{dpxdt}) and Eq. (\ref{dpydt}). This means that, in order to fully describe the electron dynamics, the two above conserved quantities are needed. This results are also valid for the operators representation of the momentum.

\section{Discussion}
Using the same method developed by Moshinsky and Quesne, and applied by Boon and Seligman, it was shown how to obtain a wave function that describes oscillations along both axes of the plane, and therefore represents a non-free electron. However, the authors referenced in \cite{boon1973canonical} are aware that this method can produce oscillations in any direction of the plane. This argument refers to the possibility of having different wave functions due to the selection of different gauges. For instance, if one were to choose the alternative Landau gauge ${{\bf A}=(0,Bx,0)}$, it would lead to a solution with oscillations along the $x$-axis instead of the $y$-axis. However, the situation with the wave function Eq.(\ref{nonfree}) turns out to be different, since it does not depend on the selection of the gauge used. Instead, it is a consequence of the conservation properties of the system and is necessary to fully describe it. In fact, due to the conservation properties Eq. (\ref{Ncons1}) and Eq.(\ref{Ncons2}), it is possible for the alternative Landau's gauge and the symmetric gauge, to find two conserved canonic momentum describing oscillations along each axis of the plane.  

\section*{Acknowledgment}
The author is thankful to Thomas Seligman for suggesting a review of his work and for the insightful discussions explaining it.

\newpage
\bibliographystyle{apsrev4-1}
\bibliography{bibliografia}

\begin{thebibliography}{6}%
\makeatletter
\providecommand \@ifxundefined [1]{%
 \@ifx{#1\undefined}
}%
\providecommand \@ifnum [1]{%
 \ifnum #1\expandafter \@firstoftwo
 \else \expandafter \@secondoftwo
 \fi
}%
\providecommand \@ifx [1]{%
 \ifx #1\expandafter \@firstoftwo
 \else \expandafter \@secondoftwo
 \fi
}%
\providecommand \natexlab [1]{#1}%
\providecommand \enquote  [1]{``#1''}%
\providecommand \bibnamefont  [1]{#1}%
\providecommand \bibfnamefont [1]{#1}%
\providecommand \citenamefont [1]{#1}%
\providecommand \href@noop [0]{\@secondoftwo}%
\providecommand \href [0]{\begingroup \@sanitize@url \@href}%
\providecommand \@href[1]{\@@startlink{#1}\@@href}%
\providecommand \@@href[1]{\endgroup#1\@@endlink}%
\providecommand \@sanitize@url [0]{\catcode `\\12\catcode `\$12\catcode
  `\&12\catcode `\#12\catcode `\^12\catcode `\_12\catcode `\%12\relax}%
\providecommand \@@startlink[1]{}%
\providecommand \@@endlink[0]{}%
\providecommand \url  [0]{\begingroup\@sanitize@url \@url }%
\providecommand \@url [1]{\endgroup\@href {#1}{\urlprefix }}%
\providecommand \urlprefix  [0]{URL }%
\providecommand \Eprint [0]{\href }%
\providecommand \doibase [0]{http://dx.doi.org/}%
\providecommand \selectlanguage [0]{\@gobble}%
\providecommand \bibinfo  [0]{\@secondoftwo}%
\providecommand \bibfield  [0]{\@secondoftwo}%
\providecommand \translation [1]{[#1]}%
\providecommand \BibitemOpen [0]{}%
\providecommand \bibitemStop [0]{}%
\providecommand \bibitemNoStop [0]{.\EOS\space}%
\providecommand \EOS [0]{\spacefactor3000\relax}%
\providecommand \BibitemShut  [1]{\csname bibitem#1\endcsname}%
\let\auto@bib@innerbib\@empty
%</preamble>
\bibitem [{\citenamefont {Boon}\ and\ \citenamefont
  {Seligman}(1973)}]{boon1973canonical}%
  \BibitemOpen
  \bibfield  {author} {\bibinfo {author} {\bibfnamefont {M.}~\bibnamefont
  {Boon}}\ and\ \bibinfo {author} {\bibfnamefont {T.}~\bibnamefont
  {Seligman}},\ }\href@noop {} {\bibfield  {journal} {\bibinfo  {journal}
  {Journal of Mathematical Physics}\ }\textbf {\bibinfo {volume} {14}},\
  \bibinfo {pages} {1224} (\bibinfo {year} {1973})}\BibitemShut {NoStop}%
\bibitem [{\citenamefont {Moshinsky}\ and\ \citenamefont
  {Quesne}(1971)}]{moshinsky1971linear}%
  \BibitemOpen
  \bibfield  {author} {\bibinfo {author} {\bibfnamefont {M.}~\bibnamefont
  {Moshinsky}}\ and\ \bibinfo {author} {\bibfnamefont {C.}~\bibnamefont
  {Quesne}},\ }\href@noop {} {\bibfield  {journal} {\bibinfo  {journal}
  {Journal of Mathematical Physics}\ }\textbf {\bibinfo {volume} {12}},\
  \bibinfo {pages} {1772} (\bibinfo {year} {1971})}\BibitemShut {NoStop}%
\bibitem [{\citenamefont {Quesne}\ and\ \citenamefont
  {Moshinsky}(1971)}]{quesne1971canonical}%
  \BibitemOpen
  \bibfield  {author} {\bibinfo {author} {\bibfnamefont {C.}~\bibnamefont
  {Quesne}}\ and\ \bibinfo {author} {\bibfnamefont {M.}~\bibnamefont
  {Moshinsky}},\ }\href@noop {} {\bibfield  {journal} {\bibinfo  {journal}
  {Journal of Mathematical Physics}\ }\textbf {\bibinfo {volume} {12}},\
  \bibinfo {pages} {1780} (\bibinfo {year} {1971})}\BibitemShut {NoStop}%
\bibitem [{\citenamefont {Moshinsky}\ \emph {et~al.}(1972)\citenamefont
  {Moshinsky}, \citenamefont {Seligman},\ and\ \citenamefont
  {Wolf}}]{moshinsky1972canonical}%
  \BibitemOpen
  \bibfield  {author} {\bibinfo {author} {\bibfnamefont {M.}~\bibnamefont
  {Moshinsky}}, \bibinfo {author} {\bibfnamefont {T.}~\bibnamefont {Seligman}},
  \ and\ \bibinfo {author} {\bibfnamefont {K.~B.}\ \bibnamefont {Wolf}},\
  }\href@noop {} {\bibfield  {journal} {\bibinfo  {journal} {Journal of
  Mathematical Physics}\ }\textbf {\bibinfo {volume} {13}},\ \bibinfo {pages}
  {901} (\bibinfo {year} {1972})}\BibitemShut {NoStop}%
\bibitem [{\citenamefont {de~la Pe{\~n}a}\ \emph {et~al.}(2021)\citenamefont
  {de~la Pe{\~n}a}, \citenamefont {Cetto},\ and\ \citenamefont
  {Vald{\'e}s-Hern{\'a}ndez}}]{de2021solution}%
  \BibitemOpen
  \bibfield  {author} {\bibinfo {author} {\bibfnamefont {L.}~\bibnamefont
  {de~la Pe{\~n}a}}, \bibinfo {author} {\bibfnamefont {A.~M.}\ \bibnamefont
  {Cetto}}, \ and\ \bibinfo {author} {\bibfnamefont {A.}~\bibnamefont
  {Vald{\'e}s-Hern{\'a}ndez}},\ }\href@noop {} {\bibfield  {journal} {\bibinfo
  {journal} {European Journal of Physics}\ }\textbf {\bibinfo {volume} {43}},\
  \bibinfo {pages} {015401} (\bibinfo {year} {2021})}\BibitemShut {NoStop}%
\bibitem [{\citenamefont {Landau}\ and\ \citenamefont
  {Lifshitz}(2013)}]{landau2013quantum}%
  \BibitemOpen
  \bibfield  {author} {\bibinfo {author} {\bibfnamefont {L.~D.}\ \bibnamefont
  {Landau}}\ and\ \bibinfo {author} {\bibfnamefont {E.~M.}\ \bibnamefont
  {Lifshitz}},\ }\href@noop {} {\emph {\bibinfo {title} {Quantum mechanics:
  non-relativistic theory}}},\ Vol.~\bibinfo {volume} {3}\ (\bibinfo
  {publisher} {Elsevier},\ \bibinfo {year} {2013})\ pp.\ \bibinfo {pages}
  {421--426}\BibitemShut {NoStop}%
\end{thebibliography}%

\end{document}